\documentclass[journal]{IEEEtran}

\usepackage{amssymb}
\usepackage{mathtools}
\usepackage{diagbox}
\usepackage{graphicx,cite,epsfig,amssymb,amsmath}
\usepackage{pifont}
\usepackage{bm}
\usepackage{color}
\usepackage{stmaryrd}        
\usepackage{standalone}
\usepackage{preview}           
\usepackage{xcolor}
\usepackage{subfigure}
\usepackage{bbding}
\usepackage{subfigure}
\usepackage[ruled,vlined]{algorithm2e}
\SetKw{Continue}{continue}
\SetKw{Break}{break}
\SetKw{And}{and}
\usepackage{algorithmic}
\usepackage{booktabs}
\usepackage{listings}
\usepackage{multirow}

\usepackage[font=footnotesize,skip=5pt]{caption}
\usepackage{titlesec}

\usepackage{hyperref}
\hypersetup{hypertex=true, 
  		    colorlinks=true,
            linkcolor=blue,
            anchorcolor=blue,
            citecolor=blue}
\UseRawInputEncoding

\usepackage[normalem]{ulem} 

\begin{document}

\title{Performance and Complexity of Sequential Decoding of PAC Codes}

\author{Mohsen~Moradi\textsuperscript{\href{https://orcid.org/0000-0001-7026-0682}{\includegraphics[scale=0.06]{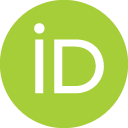}}},
        Amir~Mozammel\textsuperscript{\href{https://orcid.org/0000-0003-3474-9530}{\includegraphics[scale=0.06]{figs/ORCIDiD_icon128x128.png}}}, 
        Kangjian~Qin\textsuperscript{\href{https://orcid.org/0000-0002-9536-5453}{\includegraphics[scale=0.06]{figs/ORCIDiD_icon128x128.png}}}, and
        Erdal~Ar{\i}kan\textsuperscript{\href{https://orcid.org/0000-0001-8810-4450}{\includegraphics[scale=0.06]{figs/ORCIDiD_icon128x128.png}}},~\IEEEmembership{Fellow,~IEEE}
        \thanks{The authors are with the Department of Electrical-Electronics Engineering, Bilkent University, Ankara TR-06800, Turkey (e-mail: moradi@ee.bilkent.edu.tr, a.mozammel@ee.bilkent.edu.tr, qin@ee.bilkent.edu.tr, arikan@ee.bilkent.edu.tr).}%
}

\maketitle

\begin{abstract}
Performance and complexity of sequential decoding of polarization-adjusted convolutional (PAC) codes is studied.
In particular, a performance and computational complexity comparison of PAC codes with 5G polar codes and convolutional codes is given.
A method for bounding the complexity of sequential decoding of PAC codes is proposed.

\end{abstract}

\section{Introduction}
\IEEEPARstart{P}{olarization}-adjusted convolutional (PAC) codes are a class of liner block codes that combine ideas from channel polarization and convolutional coding \cite{arikan2009channel,arikan2019sequential}.
PAC codes show remarkably good performance at short block lengths, approaching theoretical limits in some instances.
In this paper we carry out a comparison of PAC with convolutional codes and polar codes in terms of performance and complexity.
We focus exclusively on sequential decoding of PAC codes.
PAC codes can also be decoded using fixed-complexity list-decoders as discussed in \cite{ViterboFanoListDecodingPAC2020} and \cite{VardyListDecodingPAC2020};
however, such algorithms lie outside the scope of the present paper.

We show that PAC codes under sequential decoding are capable of providing better performance than some 5G polar codes. 
We develop methods for a reasonable comparison of the computational complexity of sequential decoding of PAC codes with the list decoding of 5G polar codes. 
Finally, we consider sequential decoding under a strict limit on its search complexity so as to eliminate implementation difficulties arising from the 
variable nature of computation in sequential decoding.
Our main conclusion is that PAC codes under sequential decoding can be seen as a viable alternative to convolutional and polar codes both in terms of performance and complexity.

The rest of this paper is organized as follows.
Section \ref{sec: pac} provides a brief introduction to PAC coding and sequential decoding of PAC codes.
Section \ref{sec: conv} compares sequential decoding of PAC codes with that of ordinary convolutional codes.
Section~\ref{sec: polar} compares PAC codes with 5G polar codes.
Section~\ref{sec: fixed-complexity} studies bounded-complexity sequential decoders and quantifies the loss in performance as a result
of bounding the complexity.
Finally, Section \ref{sec: conclusion} concludes the paper with a summary and suggestions for future work.

\section{PAC Coding} \label{sec: pac}
\begin{figure}[!thb] 
\centering
	\includegraphics [width = 3.5in]{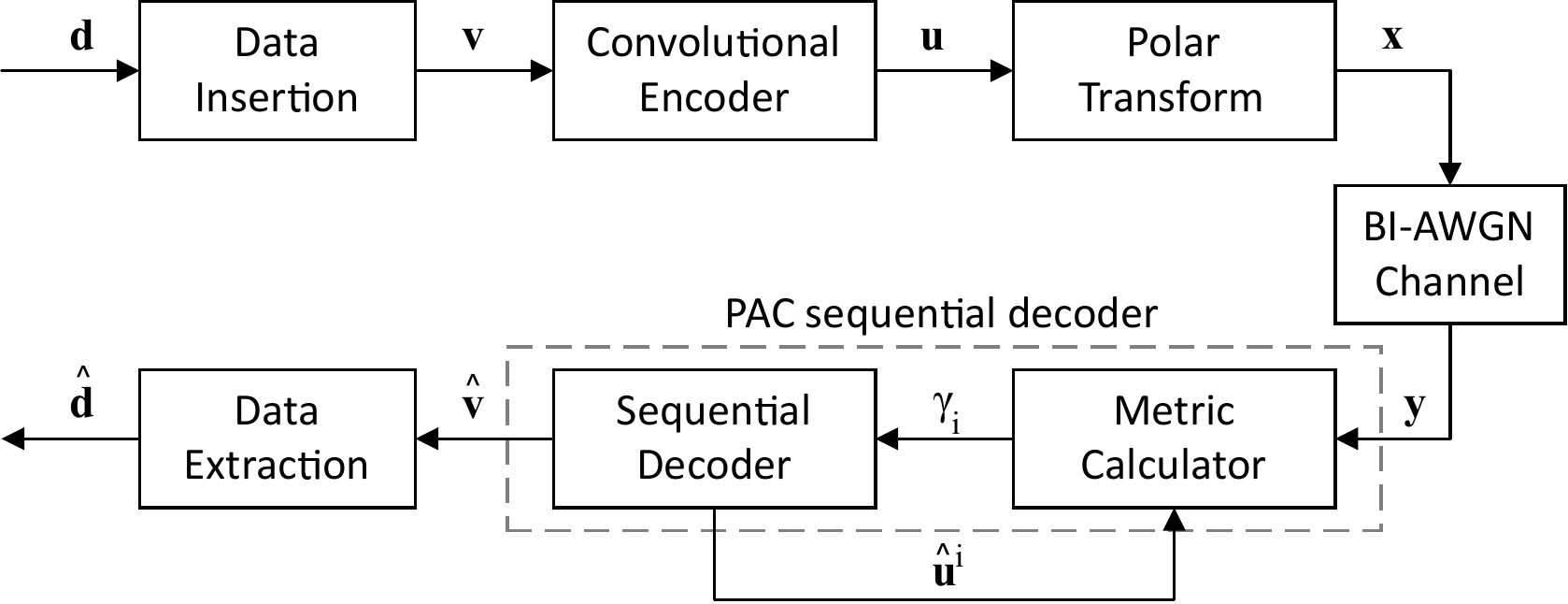}
	\caption{Flowchart of PAC coding scheme.} 
	\label{fig: flowchart}
\end{figure}

A PAC code is a linear block code over the binary field ${\mathbb F}_2=\{0,1\}$.
Fig. \ref{fig: flowchart} shows a flow chart of a PAC coding scheme with block-length $N$ and rate $R=K/N$ for some integer $1\le K\le N$.

The input to the system is a data word $\mathbf{d}=(d_1,\ldots,d_K)$ which is assumed to be chosen uniformly at random from ${\mathbb F}_2^K$.
The data insertion module maps the data word $\mathbf{d}$ to a data container $\mathbf{v}\in {\mathbb F}_2^N$ by setting $\mathbf{v}_\mathcal{A} = \mathbf{d}$ and $\mathbf{v}_{\mathcal{A}^c} = 0$,
where $\mathcal{A}$ is a data index set of size $K$. 
In this paper, the data index set $\mathcal{A}$ is chosen according to the Reed-Muller (RM) rule as explained in \cite{arikan2019sequential}.
Encoding continues by a convolution operation $\mathbf{u} = \mathbf{v}\ast \mathbf{c}$ where $\mathbf{c}=(c_0,c_1,\ldots,c_m)$ is a {\sl generator sequence}.
Next, a polar transform is applied on the convolution output to obtain the PAC codeword $\mathbf{x}=\mathbf{u}\mathbf{F}^{\otimes n}$, where $\mathbf{F}^{\otimes n}$ is the $n$th Kronecker power of a kernel matrix $\mathbf{F} = \begin{bsmallmatrix} 1 & 0\\ 1 & 1 \end{bsmallmatrix}$ and where $n = \log_2 N$.
The codeword $\mathbf{x}$ is sent through a binary-input additive white Gaussian noise (BI-AWGN) channel and a channel output $\mathbf{y}$ is received.

At the receiver side, a sequential decoder produces an estimate $\hat{\mathbf{v}}\in  {\mathbb F}_2^N$ of the data container $\mathbf{v}$ with the aid of a metric calculator.
An estimate of the data word is extracted by setting $\hat{\mathbf{d}}= \hat{\mathbf{v}}_{\mathcal A}$. 
The main performance metric for the system is the frame error rate (FER), defined as the probability $\text{P}(\hat{\mathbf{d}}\neq \mathbf{d})$. 

To study the performance of PAC codes, we implemented a sequential decoder as described in \cite{arikan2019sequential}.
Specifically, we implemented a Fano decoder \cite{Fano1963} with a metric of the form 
\begin{equation*}
\Gamma(\hat{\mathbf{u}}^i;\mathbf{y}) = \log_2\frac{\text{P}(\mathbf{y}|\hat{\mathbf{u}}^i)}{\text{P}(\mathbf{y})}-\sum_{j=1}^i b_j,
\end{equation*}
where $\hat{\mathbf{u}}^{i}$ is a candidate path at level $i$ of a search tree corresponding to the PAC code and  
$b_j = \rho$ for $j\in {\mathcal A}$ and $b_j = 0$ for $j\notin {\mathcal A}$ for some constant $\rho$.
This metric is calculated incrementally as a sum of branch metrics of the form
\begin{equation*}\label{BranchMetric}
\begin{aligned}
\gamma(\hat{u}_j;\mathbf{y},\hat{\mathbf{u}}^{j-1}) & = \Gamma(\hat{\mathbf{u}}^j;\mathbf{y}) - \Gamma(\hat{\mathbf{u}}^{j-1};\mathbf{y})\\
&= \log_2\frac{\text{P}(\mathbf{y},\hat{\mathbf{u}}^{j-1}|\hat{u}_j)}{\text{P}(\mathbf{y},\hat{\mathbf{u}}^{j-1})}-b_j.
\end{aligned}
\end{equation*}

In PAC code simulations presented below, we used a version of the Fano decoder as described in Section 6.9 of \cite{r._g._gallager_information_1968}.
The bias parameter $\rho$ for the Fano metric and the threshold spacing $\Delta$ for the Fano decoder are specified in the related section.

Sequential decoder carries out a search for the correct path in the code tree corresponding to the PAC code.
The search complexity is a random variable that depends, among other things, on the severity of noise. 
In the following, we will measure the complexity of sequential decoding by a random variable $Z$, which counts the number of nodes in the decoding
tree that are visited by the Fano algorithm during a decoding session. 
Since the Fano algorithm has a backtracking feature, it may visit some nodes more than once, and $Z$ counts each such visit.
The variable $Z$ only counts the nodes visited by a forward move and ignores lateral or backward moves \cite[p.~273]{r._g._gallager_information_1968}.
(A version of sequential decoding, called the stack algorithm \cite{zigangirov1966some,jelinek1969fast}, visits each node at most once but it uses more memory and requires sorting of nodes in a stack.)
We will be primarily interested in the expectation $E(Z)$, which we will call {\sl average number of visits} (ANV) and estimate it by computing the empirical mean of 
$Z$ over a sufficiently large number of simulation runs.

\section{Comparison with Convolutional Codes} \label{sec: conv}
\begin{figure}[!h] 
\centering
	\includegraphics [width = 3.5in]{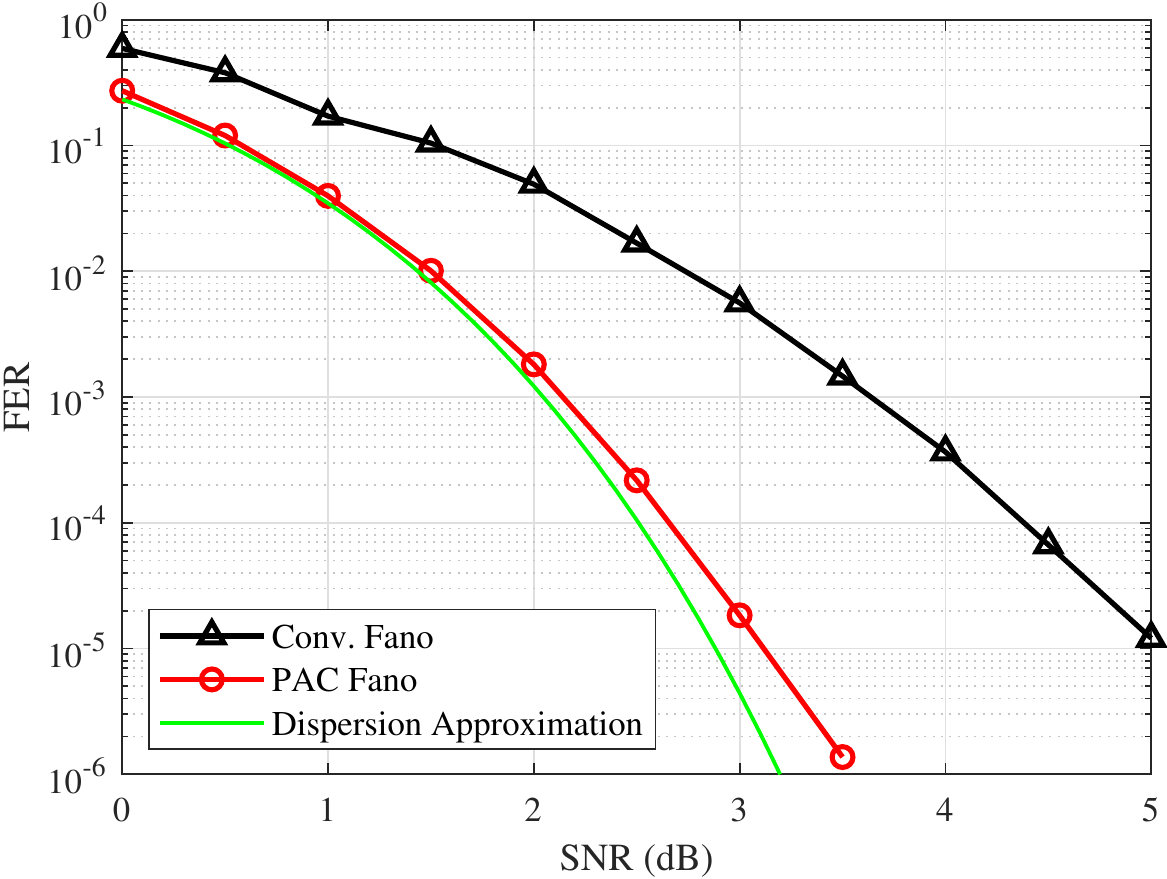}
	\caption{FER performance of PAC and convolutional codes.} 
	\label{fig: PAC vs Conv}
\end{figure}

In this section, we compare sequential decoding of PAC codes with that of ordinary convolutional codes.
Fig. \ref{fig: PAC vs Conv} shows the result of such a comparison from the viewpoint of FER performance,
with a favorable outcome for PAC codes. Also shown in Fig. \ref{fig: PAC vs Conv} is the dispersion approximation, 
which approximates the best achievable performance by any code of a given length and rate \cite{polyanskiy_channel_2010}.
PAC code shows a near-optimal performance over a range of SNR ($E_s/N_0$ dB) values.

The convolutional code in Fig.~\ref{fig: PAC vs Conv} uses a pair of generator sequences $\mathbf{c} = (133,171)$ (in octal notation) and encodes $K=64$ message bits into $N=140$ codeword bits (with 12 termination bits),
for an effective coding rate of $R = 64/140 = 0.457$. 
The corresponding parameters for the PAC code in Fig.~\ref{fig: PAC vs Conv} are $\mathbf{c} = 133$, $K = 59$, $N = 128$, and $R=0.461$. 
The bias and threshold spacing parameters in sequential decoding of the PAC code are fixed as $\rho = 1.35$ and $\Delta = 2$.

\begin{figure}[!h] 
\centering
	\includegraphics [width = 3.5in]{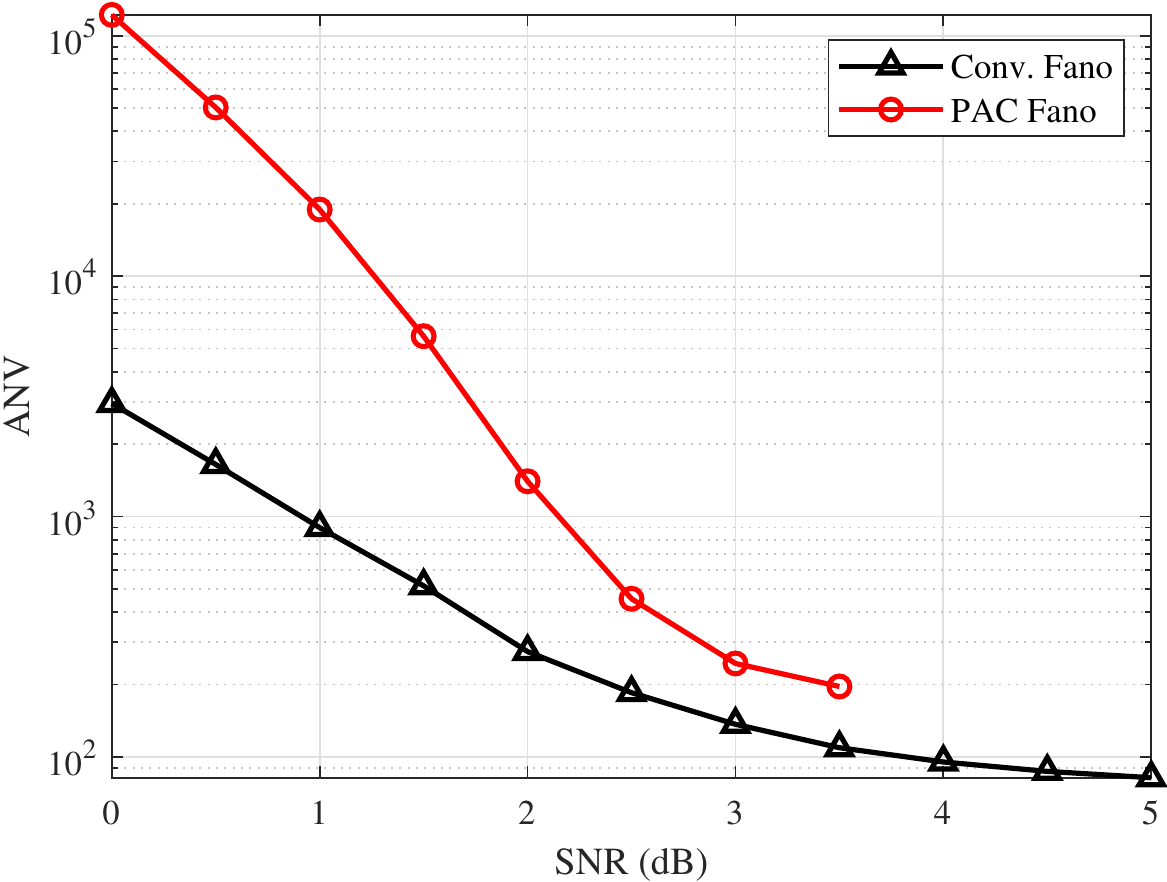}
	\caption{Average number of visits for PAC and convolutional Fano decoders.} 
	\label{fig: NV vs SNR}
\end{figure}

Next, we turn to a comparison of computational complexity between the two codes. 
Fig. \ref{fig: NV vs SNR} displays the ANV for sequential decoding of the two codes. 
At low SNR points, the PAC Fano decoder has a significantly larger average complexity as measured by the ANV, but as SNR increases, the complexity gap between the two sequential decoders decreases.

A fundamental result on sequential decoding of ordinary convolutional codes states that $Z$ has a heavy-tailed Pareto distribution \cite{jacobs_lower_1967}. 
Further, for rates $R$ above a cutoff rate $R_0$, the average computation $E(Z)$ increases exponentially in $N(R-R_0)$ \cite{jacobs_lower_1967, arikan_inequality_1996}.
If we ignore the termination, the convolutional code here operates at rate $1/2$ and we have $R_0=1/2$ at 2.46 dB SNR.
Simulation results show a significant rise in computation complexity at SNRs below 2.5 dB, as predicted by theory. 
We also see that the ANV for the sequential decoding of the PAC code experiences a sharp rise at SNR values below 2.5 dB.
So, the PAC codes also suffer from a cutoff rate phenomenon. This is not unexpected in view of a data processing theorem \cite[p.~149-150]{r._g._gallager_information_1968}
that states that cutoff rate cannot be increased by pre- and post-processing operations on a channel.

We should point out that it is actually possible to change the performance versus complexity tradeoff in sequential decoding of PAC codes by changing the data index set $\mathcal{A}$.
For example, in experiments (not reported here) we observed that if $\mathcal{A}$ is chosen in accordance with a polar code design rule, the ANV decreases; however, the reduction in ANV takes place at the expense of a worsening of the FER performance.

\section{Comparison with 5G Polar Codes} \label{sec: polar}

\begin{figure}[!htb]
    \centering
	\includegraphics [width = 3.5in]{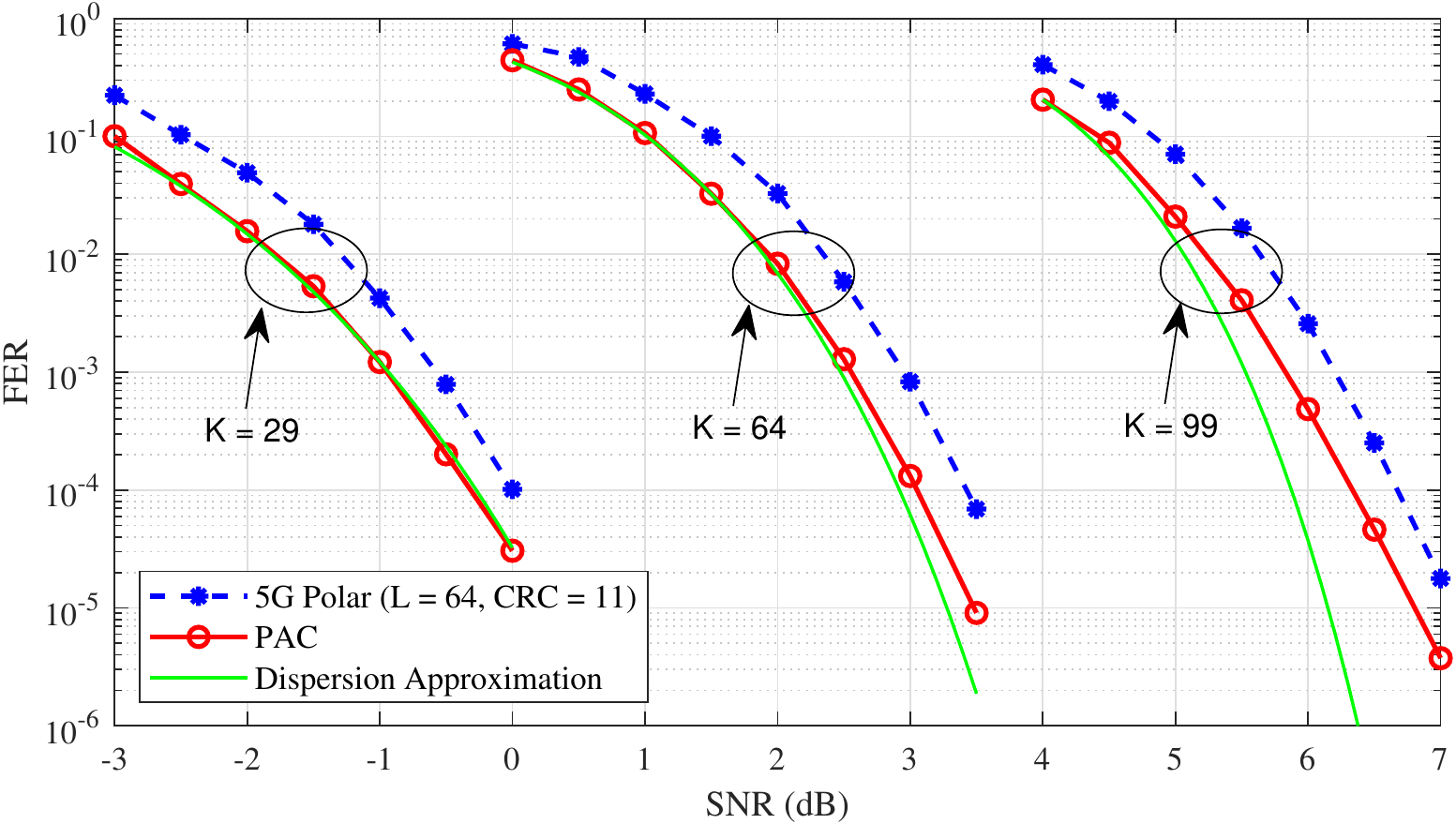}
	\caption{FER performance of PAC and 5G polar codes.} 
	\label{fig: PAC vs 5G}
\end{figure}

Fig. \ref{fig: PAC vs 5G} compares the FER performance of PAC codes and 5G polar codes \cite{3GPP} for three different code rates with $K \in \{29,64,99\}$ and $N = 128$.
The 5G polar codes had a CRC of length 11. To decode the 5G polar codes, we used a CRC-aided successive cancellation list (CA-SCL) decoder from MATLAB\textsuperscript{\textregistered} 5G Toolbox\textsuperscript{TM} with a list size of $L = 64$. All PAC codes in the comparison were constructed using the RM design rule. 
Table \ref{tab: Param} lists the generator sequences $\mathbf{c}$ for the PAC encoder and the bias $\rho$ and threshold spacing $\Delta$ parameters for the Fano decoder.
The generator sequence $\mathbf{c} = 3211$ is obtained by an ad-hoc search method with the goal of maximizing FER performance.
As Fig. \ref{fig: PAC vs 5G} shows, for the three specific cases considered here, PAC codes perform better than 5G polar codes.

\begin{table}[!ht]
\begin{center}
\caption{PAC encoder and decoder parameters.}
\begin{tabular}{|c|c|c|c|c|}
\hline
$N$ & $K$ & $\mathbf{c}$ & $\rho$ & $\Delta$   \\ \hline \hline
128 & 29 & 3211 & 1.4    & 2         \\ \hline
128 & 64 & 133 & 1.35   & 2        \\ \hline 
128 & 99 & 133 & 1.14   & 2         \\ \hline
\end{tabular}
\label{tab: Param}
\end{center}
\end{table}

The ANV figures for the three PAC codes are listed in Table \ref{tab: ANV}.
Sequential decoder has a built-in mechanism for adjusting search complexity in response to severity of channel noise, which explains why
complexity becomes smaller as the SNR is increased. The SNR values corresponding to $R_0$ being equal to $29/128$, $64/128$, and $99/128$
are -1.63, 2.46, and 5.49 dB, respectively. The computation in sequential decoding of PAC codes sees a sharp rise as the SNR crosses 
the cutoff rate thresholds from above.

\begin{table}[!ht]
\begin{center}
\caption{Average number of visits per codeword for PAC Fano decoder.}
\begin{tabular}{|c|c||c|c||c|c|}
\hline
\multicolumn{2}{|c|}{$K = 29$} & \multicolumn{2}{c|}{$K = 64$} & \multicolumn{2}{c|}{$K = 99$} \\ \hline
SNR (dB)         & AVN         & SNR (dB)       & AVN          & SNR (dB)        & AVN         \\ \hline \hline
-3               & 1301        & 0              & 219963       & 4               & 6569        \\ \hline
-2.5             & 770         & 0.5            & 87166        & 4.5             & 2570        \\ \hline
-2               & 509         & 1              & 32237        & 5               & 826         \\ \hline
-1.5             & 344         & 1.5            & 9369         & 5.5             & 401         \\ \hline
-1               & 256         & 2              & 2749         & 6               & 303         \\ \hline
-0.5             & 213         & 2.5            & 736          & 6.5             & 279         \\ \hline
0                & 191         & 3              & 368          & 7               & 270         \\ \hline
                 &             & 3.5            & 287          &                 &             \\ \hline
\end{tabular}
\label{tab: ANV}
\end{center}
\end{table}

We will measure the complexity of a CA-SCL decoder for polar codes by the total number of nodes that the decoder processes. 
A list decoder begins with an empty list and builds up a list of size $L$ after $\log_2 L$ branching points in the code tree for polar codes.
Thereafter, the list decoder examines $2L$ candidate nodes at each branching point. We take the number of branching points in the code tree to be $K+C$ where $C$ is the number of parity or CRC bits.
We ignore the lower complexity of the first $\log_2 L$ steps and approximate the total number of nodes inspected by a list-of-$L$ decoder by $2L(K+C)$. 
Note that this complexity estimate excludes the sorting complexity of a list decoder, which actually may dominate the complexity for a large list size, such as $L=64$.

For the 5G polar codes here, we have $L=64$, $C=11$, and $K\in \{29,64,99\}$. 
Accordingly, we estimate the CA-SCL decoder complexity as 5120 for $K=29$, 9600 for $K=64$, and 14080 for $K=99$.
Comparing these 5G polar complexity figures with the PAC complexity figures in Table~\ref{tab: ANV}, we see that, for message lengths $K=29$ and $K=99$, 
decoding of PAC codes is significantly less complex than that of 5G polar codes on average. 
For $K=64$, the decoding complexity of the PAC code is smaller than that of the 5G polar code at SNR values greater than 1.5 dB.
Based on this comparison, we conclude that the PAC decoder has a complexity comparable to that of CA-SCL decoder.

\section{Bounded-complexity PAC Fano Decoder}\label{sec: fixed-complexity}

Sequential decoding is a variable-complexity decoding algorithm. In many applications it is desirable to have a fixed- or bounded-complexity decoding method.
In this section, we consider imposing a strict limit on the complexity of sequential decoding and study its performance under such a constraint.
For this we need to study the distribution of the complexity measure $Z$ rather than ANV, which corresponds to the mean complexity $E(Z)$. 

To this end, we performed a simulation study of sequential decoding of the above $(128,64)$ PAC code and recorded the observed $Z$ values. We
discarded the $Z$ values when a decoding error occurred and kept the $Z$ values in correct decodings.
The results are shown in Table \ref{tab: VH} for SNR equal to 2.5 dB and 3.0 dB.
The heavy-tailed nature of distribution of $Z$ is confirmed by the data in Table~\ref{tab: VH}. A small fraction of decoding instances require a very large decoding complexity.
Motivated by this we consider stopping the sequential decoder and declaring a decoder error when $Z$ exceeds a parameter $Z_\text{max}$.

\begin{table}[!ht]
\begin{center}
\caption{Frequency distribution of number of visits $Z$}
\begin{tabular}{|c||c||c|}
\hline
Number of visits         & 2.5 dB (\%) & 3.0 dB (\%) \\ \hline \hline
$Z \leq 2^{10}$          & 92.8475     & 98.1415   \\ \hline
$2^{10} < Z \leq 2^{11}$ & 3.6951      & 1.1643    \\ \hline
$2^{11} < Z \leq 2^{12}$ & 1.7146      & 0.4120    \\ \hline
$2^{12} < Z \leq 2^{13}$ & 0.8982      & 0.1707    \\ \hline
$2^{13} < Z \leq 2^{14}$ & 0.4551      & 0.0679    \\ \hline
$2^{14} < Z \leq 2^{15}$ & 0.2369      & 0.0279    \\ \hline
$2^{15} < Z \leq 2^{16}$ & 0.1060      & 0.0115    \\ \hline
$2^{16} < Z \leq 2^{17}$ & 0.0388      & 0.0032    \\ \hline
$2^{17} < Z \leq 2^{18}$ & 0.0073      & 0.0009    \\ \hline
$2^{18} < Z$             & 0.0005      & 0.0001    \\ \hline
\end{tabular}
\label{tab: VH}
\end{center}
\end{table}

\begin{figure}[!thb] 
\centering
	\includegraphics [width = 3.5in]{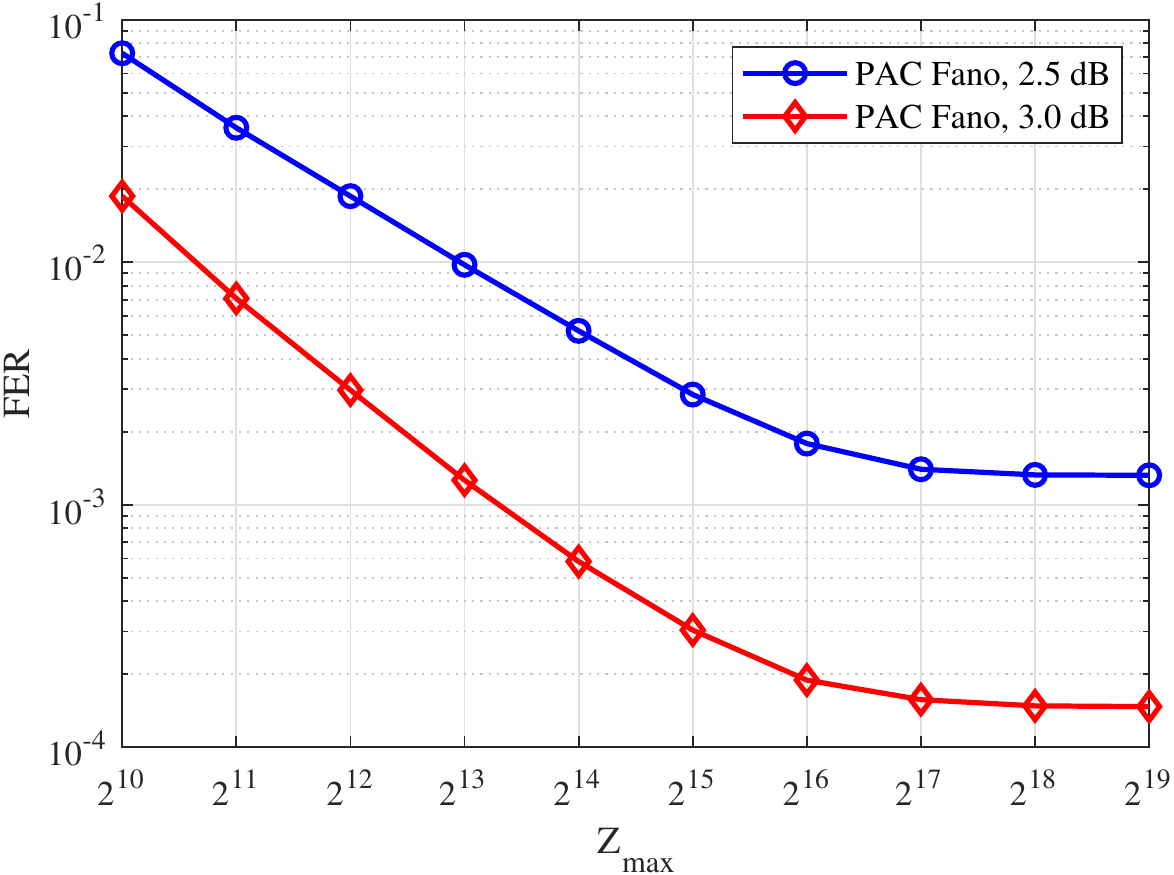}
	\caption{FER v. $Z_{\text{max}}$ for Fano decoder.} 
	\label{fig: FER_vs_Zmax}
\end{figure}

\begin{figure}[!tbh] 
\centering
	\includegraphics [width = 3.5in]{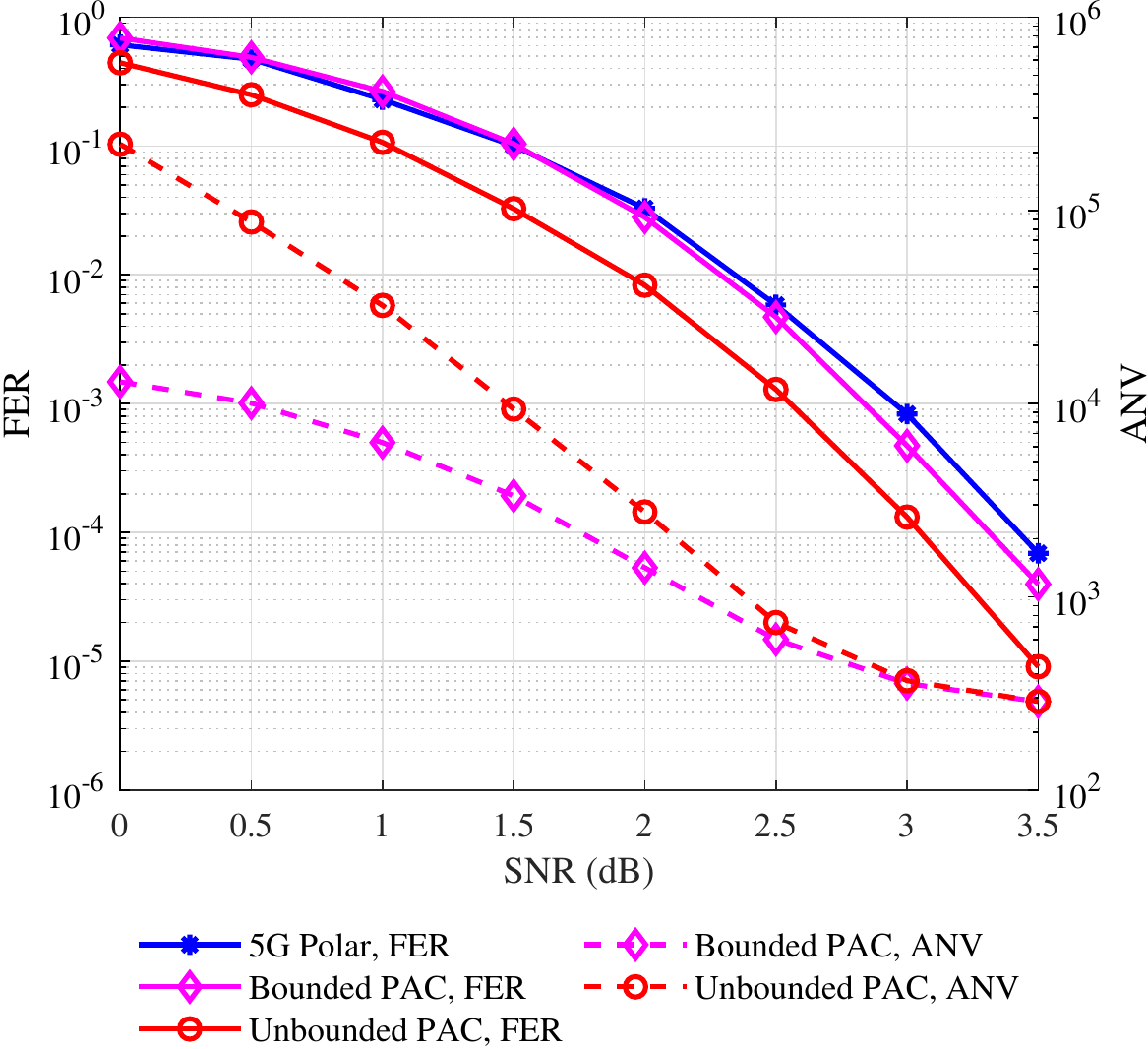}
	\caption{Performance and complexity of Fano decoder with $Z_{\text{max}} = 16384$.} 
	\label{fig: lim_pac_perf}
\end{figure}

Fig. \ref{fig: FER_vs_Zmax} displays the trade-off between FER and $Z_{\text{max}}$ at two different SNR values. 
Such figures prove to be useful in deciding the value of $Z_\text{max}$.
For example, suppose the goal is to match the FER performance of the (128,64) 5G polar code at 2.5 dB SNR.
From Fig. \ref{fig: PAC vs 5G}, we read the target FER value as $6 \times 10^{-3}$.
Fig. \ref{fig: FER_vs_Zmax} indicates that this FER target is achievable with $Z_{\text{max}} = 2^{14}=16384$.
Fig. \ref{fig: lim_pac_perf} confirms that the resulting bounded-complexity Fano decoder does indeed achieve the target FER at 2.5 dB for a PAC$(128,64)$ code.
In comparison, the CA-SCL decoder for the 5G polar code in Fig.~\ref{fig: PAC vs 5G} has a complexity of 9600 node visits.

\section{Conclusion}\label{sec: conclusion}
We showed that PAC codes under sequential decoding have the potential to improve the performance of 5G polar codes.
The main drawback of sequential decoding is its variable complexity. We addressed this problem by introducing a bound
on the complexity of sequential decoding. 
In this connection, one may consider limiting the total number of node visits by introducing per-level limits on the number of node visits to reduce
the computational complexity still further without significantly degrading FER performance.

We considered PAC codes based only on the RM design rule for selection of the data index set $\mathcal{A}$.
The RM design rule maximizes the minimum distance $d_\text{min}$ of PAC codes as shown in \cite{li2019pre}, which
is important for FER performance at high SNR. 
On the other hand, the selection of $\mathcal{A}$ in accordance with the RM rule is not exactly compatible with the channel 
polarization created by the polar transform, which results in high computational complexity.
Another subject for future study is to compare alternative design rules for $\mathcal{A}$ with respect to performance and complexity.

The connection sequence $\mathbf{c}$ also has a significant impact on the performance of PAC codes through its effect on the multiplicity 
$A_{d_\text{min}}$ of minimum-distance codewords. 
The generator sequence $\mathbf{c} = 3211$ used in the PAC code with dimension $K = 29$ was obtained by an ad-hoc search method with the goal of minimizing
$A_{d_\text{min}}$.
The choice $\mathbf{c} = 3211$ improved the FER performance significantly compared to our default generator sequence $\mathbf{c} = 133$. 
Identifying good generator sequences in connection with other design parameters of PAC codes remains another subject for future study.

\ifCLASSOPTIONcaptionsoff
  \newpage
\fi

\bibliographystyle{IEEEtran}


\end{document}